# A quark-meson coupling model for nuclear and neutron matter

K. Saito*

Physics Division, Tohoku College of Pharmacy

Sendai 981, Japan

and

A. W. Thomas[†]

Department of Physics and Mathematical Physics

University of Adelaide, South Australia, 5005, Australia

March 18, 1994

**Abstract**

An explicit quark model, based on a mean field description of non-overlapping nucleon bags bound by the self-consistent exchange of $\sigma$, $\omega$ and $\rho$ mesons, is used to investigate the properties of both nuclear and neutron matter. We establish a clear understanding of the relationship between this model, which incorporates the internal structure of the nucleon, and QHD. Finally, we use the model to study the density dependence of the quark condensate in-medium.

---

*Correspondence to: Dr. K. Saito, e-mail: ksaito@nucl.phys.tohoku.ac.jp

[†]e-mail: athomas@physics.adelaide.edu.au

Recently there has been considerable interest in relativistic calculations of infinite nuclear matter as well as dense neutron matter. A relativistic treatment is, of course, essential if one aims to deal with the properties of dense matter, including the equation of state (EOS). The simplest relativistic model for hadronic matter is the Walecka model [1] (often called Quantum Hadrodynamics, *i.e.*, QHD-I [2]), which consists of *structureless* nucleons interacting through the exchange of the $\sigma$ meson and the time component of the $\omega$ meson in the mean-field approximation (MFA). Later Serot and Walecka extended the model to incorporate the isovector mesons, $\pi$ and $\rho$ (QHD-II [2, 3]) and used it to discuss systems like neutron stars with $N \neq Z$. QHD has proven to be a very powerful tool for the treatment of a wide range of nuclear phenomena [2]. However, it should be noted that a strong excitation of nucleon-antinucleon pairs due to the meson fields is somewhat unlikely from the point of view of the quark model because the meson fields create only one quark-antiquark pair at a time. An extension of such a model to include the structure of the nucleon may provide some new insight into the properties of hadronic matter.

A quark-meson coupling (QMC) mechanism for the saturation of nuclear matter was initially proposed by Guichon [4]. In his simple model, nuclear matter consists of non-overlapping nucleon bags bound by the self-consistent exchange of scalar ($\sigma$) and vector ($\omega$) mesons in the MFA. This model has been generalised to include nucleon Fermi motion as well as the center-of-mass (c.m.) correction to the bag energy [5]. Using this model, it has proven possible to investigate both the change of nucleon properties in nuclear matter and nuclear structure functions [6, 7]. (An alternative version of this model, based on the color dielectric soliton model, has been developed by Naar and Birse [8], who also applied it to deep-inelastic scattering.) To treat asymmetric nuclear matter (*i.e.*, $N \neq Z$) the model requires some extension. In this work we show that it is possible to include the contribution of the isovector, vector meson, $\rho$, in the study of such systems. We shall see that it is very interesting to compare this new model with QHD-II and to investigate the properties of hadronic matter in terms of quark degrees of freedom.

Let the mean-field values for the $\sigma$, $\omega$ (the time component) and $\rho$ (the time component in the third direction of isospin) fields, in uniformly distributed nuclear matter with $N \neq Z$, be $\bar{\sigma}$, $\bar{\omega}$ and $\bar{b}$, respectively. The nucleon is described by the static spherical MIT bag [9] in which quarks interact with those mean fields. The Dirac equation for a quark field, $\psi_q$, in a bag is then given by[1]

$$(i\gamma \cdot \partial - (m_q - V_\sigma) - \gamma^0(V_\omega + \frac{1}{2}\tau_q^z V_\rho))\psi_q = 0, \tag{1}$$

where $V_\sigma = g_\sigma^q \bar{\sigma}$, $V_\omega = g_\omega^q \bar{\omega}$ and $V_\rho = g_\rho^q \bar{b}$ with the quark-meson coupling constants, $g_\sigma^q$, $g_\omega^q$ and $g_\rho^q$. The bare quark mass is denoted by $m_q$ and $\tau_q^z$ is the third component of the Pauli matrix. (Here we deal with u and d quarks (q=u or d) only.) The normalized, ground state for a quark in the nucleon is then given by

$$\psi_q(\vec{r}, t) = \mathcal{N} e^{-i\epsilon_q t/R} \begin{pmatrix} j_0(xr/R) \\ i\beta_q \vec{\sigma} \cdot \hat{r} j_1(xr/R) \end{pmatrix} \frac{\chi_q}{\sqrt{4\pi}}, \tag{2}$$

where

$$\epsilon_q = \Omega_q + R(V_\omega \pm \frac{1}{2}V_\rho), \text{ for } \begin{pmatrix} u \\ d \end{pmatrix} \text{ quark} \tag{3}$$

$$\mathcal{N}^{-2} = 2R^3 j_0^2(x)[\Omega_q(\Omega_q - 1) + Rm_q^\star/2]/x^2, \tag{4}$$

$$\beta_q = \sqrt{\frac{\Omega_q - Rm_q^\star}{\Omega_q + Rm_q^\star}}, \tag{5}$$

with $\Omega_q = \sqrt{x^2 + (Rm_q^\star)^2}$ and $\chi_q$ the quark spinor. The effective quark mass, $m_q^\star$, is defined by

$$m_q^\star = m_q - V_\sigma = m_q - g_\sigma^q \bar{\sigma}. \tag{6}$$

The boundary condition at the surface provides the equation for the eigenvalue $x$:

$$j_0(x) = \beta_q j_1(x). \tag{7}$$

Using the SU(6) spin-flavor nucleon wave function, $\left| \begin{pmatrix} p \\ n \end{pmatrix} \right\rangle$, the nucleon energy is given by

$$E_N = \left\langle \begin{pmatrix} p \\ n \end{pmatrix} \middle| E_{bag} + 3V_\omega + \frac{1}{2}V_\rho \sum_q \tau_q^z \middle| \begin{pmatrix} p \\ n \end{pmatrix} \right\rangle, \tag{8}$$

$$= E_{bag} + 3V_\omega \pm \frac{1}{2}V_\rho, \text{ for } \begin{pmatrix} \text{proton} \\ \text{neutron} \end{pmatrix}. \tag{9}$$

---

[1]The sign of the $\sigma$ field value in this paper is opposite to that in Ref.[7] in order to simplify the comparison with QHD.

Here the bag energy is

$$E_{bag} = \frac{\sum_q \Omega_q - z_0}{R} + \frac{4}{3}\pi B R^3, \qquad (10)$$

with $B$ the bag constant and $z_0$ the usual parameter which accounts for zero-point motion.

In order to correct for spurious c.m. motion in the bag [10] the mass of the nucleon at rest is taken to be

$$M_N = \sqrt{E_{bag}^2 - \langle p_{cm}^2 \rangle}, \qquad (11)$$

where $\langle p_{cm}^2 \rangle = \sum_q \langle p_q^2 \rangle$ and $\langle p_q^2 \rangle$ is the expectation value of the quark momentum squared, $(x/R)^2$. Inside nuclear matter the effective nucleon mass, $M_N^\star$, is also given by minimizing eq.(11) with respect to $R$. To test the sensitivity of our results to the radius of the free nucleon, $R_0$, we vary $B$ and $z_0$ to fix $R_0 (= 0.6, 0.8, 1.0 fm)$ and the free nucleon mass ($M_N = 939$ MeV). The values of $B$ and $z_0$ for several choices of radius for the free nucleon

Table 1: $B^{1/4}$ and $z_0$ for some bag radii ($m_q = 0$ MeV).

| $R_0(fm)$ | 0.6 | 0.8 | 1.0 |
|---|---|---|---|
| $B^{1/4}$(MeV) | 188.1 | 157.5 | 136.3 |
| $z_0$ | 2.030 | 1.628 | 1.153 |

are summarized in table 1.

The Pauli principle induces Fermi motion of the nucleon. Since the energy of a moving bag with momentum $\vec{k}$ is

$$\varepsilon(\vec{k}) = \sqrt{M_N^{\star 2} + \vec{k}^2} + 3V_\omega \pm \frac{1}{2}V_\rho, \text{ for } \begin{pmatrix} \text{proton} \\ \text{neutron} \end{pmatrix} \qquad (12)$$

the total energy per nucleon at the nuclear density, $\rho_B$, is given by

$$E_{tot} = \frac{2}{\rho_B (2\pi)^3}(\int^{k_{F_p}} + \int^{k_{F_n}})d\vec{k}\sqrt{M_N^{\star 2} + \vec{k}^2} + 3V_\omega + \frac{V_\rho}{2}\left(\frac{\rho_3}{\rho_B}\right) + \frac{1}{2\rho_B}(m_\sigma^2 \bar{\sigma}^2 - m_\omega^2 \bar{\omega}^2 - m_\rho^2 \bar{b}^2). \qquad (13)$$

Here $\rho_3$ is the difference between the proton and neutron densities, $\rho_p - \rho_n$, and $k_{F_p}$ and $k_{F_n}$ are the Fermi momenta for protons and neutrons, respectively:

$$\rho_p = \frac{1}{3\pi^3} k_{F_p}^3, \quad \rho_n = \frac{1}{3\pi^3} k_{F_n}^3 \quad \text{and} \quad \rho_B = \rho_p + \rho_n. \tag{14}$$

The $\omega$ field created by uniformly distributed nucleons is determined by baryon number conservation to be

$$\bar{\omega} = \frac{V_\omega}{g_\omega^q} = \frac{3 g_\omega^q \rho_B}{m_\omega^2} = \frac{g_\omega \rho_B}{m_\omega^2}, \tag{15}$$

(where we define a new coupling constant $g_\omega = 3g_\omega^q$) while the $\sigma$ and $\rho$ mean-fields are given by the thermodynamic conditions [2]:

$$\left( \frac{\partial E_{tot}}{\partial \bar{\sigma}} \right)_{R,\rho_B} = 0 \quad \text{and} \quad \left( \frac{\partial E_{tot}}{\partial \bar{b}} \right)_{R,\rho_B} = 0. \tag{16}$$

Since the $\rho$ field value is expressed by

$$\bar{b} = \frac{g_\rho}{2 m_\rho^2} \rho_3, \tag{17}$$

where we take $g_\rho = g_\rho^q$, the total energy per nucleon can be rewritten as

$$E_{tot} = \frac{2}{\rho_B (2\pi)^3} \left( \int^{k_{F_p}} + \int^{k_{F_n}} \right) d\vec{k} \sqrt{M_N^{\star 2} + \vec{k}^2} + \frac{g_\omega^2}{2m_\omega^2} \rho_B + \frac{m_\sigma^2}{2\rho_B} \bar{\sigma}^2 + \frac{g_\rho^2}{8 m_\rho^2 \rho_B} \rho_3^2. \tag{18}$$

Therefore, using eq.(16), the value of the $\sigma$ field is given by

$$\bar{\sigma} = -\frac{1}{m_\sigma^2} \frac{2}{(2\pi)^3} \left( \int^{k_{F_p}} + \int^{k_{F_n}} \right) d\vec{k} \frac{M_N^\star}{\sqrt{M_N^{\star 2} + \vec{k}^2}} \times \left( \frac{\partial M_N^\star}{\partial \bar{\sigma}} \right)_R. \tag{19}$$

We should note that the expression for the total energy, eq.(18), is identical to that of QHD-II. The effect of the internal quark structure enters entirely through the effective nucleon mass, eqs.(10) and (11), and the self-consistency condition (SCC) for the $\sigma$ field, eq.(19).

Let us consider the SCC further. The effective quark mass, $m_q^\star$, is given by eq.(6). If the nucleon were simply made of three massive constituent quarks the nucleon mass in vacuum and that in matter would satisfy

$$M_N \approx 3 m_q \quad \text{and} \quad M_N^\star \approx 3 m_q^\star, \tag{20}$$

and hence, using eq.(6), the effective nucleon mass could be

$$M_N^\star = M_N - g_\sigma \bar{\sigma}, \tag{21}$$

where we defined $g_\sigma = 3g_\sigma^q$. Since one finds

$$\left(\frac{\partial M_N^\star}{\partial \bar{\sigma}}\right) = -g_\sigma, \tag{22}$$

from eq.(21), the SCC gives

$$g_\sigma \bar{\sigma} = 3g_\sigma^q \bar{\sigma} \tag{23}$$

$$= \frac{g_\sigma^2}{m_\sigma^2} \frac{2}{(2\pi)^3} \left(\int^{k_{F_p}} + \int^{k_{F_n}}\right) d\vec{k} \frac{M_N^\star}{\sqrt{M_N^{\star 2} + \vec{k}^2}}. \tag{24}$$

Together with eq.(21), we then find the SCC in the heavy quark mass limit:

$$M_N^\star = M_N - g_\sigma \bar{\sigma} \tag{25}$$

$$= M_N - \frac{g_\sigma^2}{m_\sigma^2} \frac{2}{(2\pi)^3} \left(\int^{k_{F_p}} + \int^{k_{F_n}}\right) d\vec{k} \frac{M_N^\star}{\sqrt{M_N^{\star 2} + \vec{k}^2}}. \tag{26}$$

This is exactly the SCC for QHD-II [2].

However, in the more realistic case of light quarks the internal structure of the nucleon significantly alters the SCC. In particular, the u and d quarks in the bag model, like current quarks, have very small masses and must therefore be treated in a fully relativistic manner. Using eqs.(10) and (11), one finds

$$\left(\frac{\partial M_N^\star}{\partial \bar{\sigma}}\right)_R = -g_\sigma \times \left(\frac{E_{bag}}{M_N^\star}\right) \left[(1 - \frac{\Omega_q}{E_{bag} R}) S(\bar{\sigma}) - \frac{V_\sigma}{E_{bag}}\right] \tag{27}$$

$$\equiv -g_\sigma \times C(\bar{\sigma}), \tag{28}$$

where $S(\bar{\sigma})$ is the scalar density of the nucleon bag in matter:

$$S(\bar{\sigma}) = \int d\vec{r} \bar{\psi}_q \psi_q = \frac{\Omega_q/2 + Rm_q^\star(\Omega_q - 1)}{\Omega_q(\Omega_q - 1) + Rm_q^\star/2}. \tag{29}$$

Hence we find the SCC for the $\sigma$ field in the case where the nucleon has the quark structure:

$$g_\sigma \bar{\sigma} = \frac{g_\sigma^2}{m_\sigma^2} \frac{2}{(2\pi)^3} C(\bar{\sigma}) \left(\int^{k_{F_p}} + \int^{k_{F_n}}\right) d\vec{k} \frac{M_N^\star}{\sqrt{M_N^{\star 2} + \vec{k}^2}}. \tag{30}$$

Clearly the effect of the internal, quark structure is completely absorbed into the scalar density factor, $C(\bar{\sigma})$. Thus, we see that it is possible to unify the present QMC model and QHD in the MFA: the energy per nucleon and the SCC for the $\sigma$ field are, respectively, given by eq.(18) and eq.(19) with

$$\left(\frac{\partial M_N^\star}{\partial \bar{\sigma}}\right)_R = -g_\sigma \times \begin{pmatrix} 1 \\ C(\bar{\sigma}) \end{pmatrix} \text{ for } \begin{pmatrix} \text{QHD} \\ \text{QMC} \end{pmatrix}. \qquad (31)$$

The scalar density factor is shown in fig.1 as a function of the scalar field strength for three bag radii. We see that it is much smaller than unity and that the dependence on the bag radius is not strong.

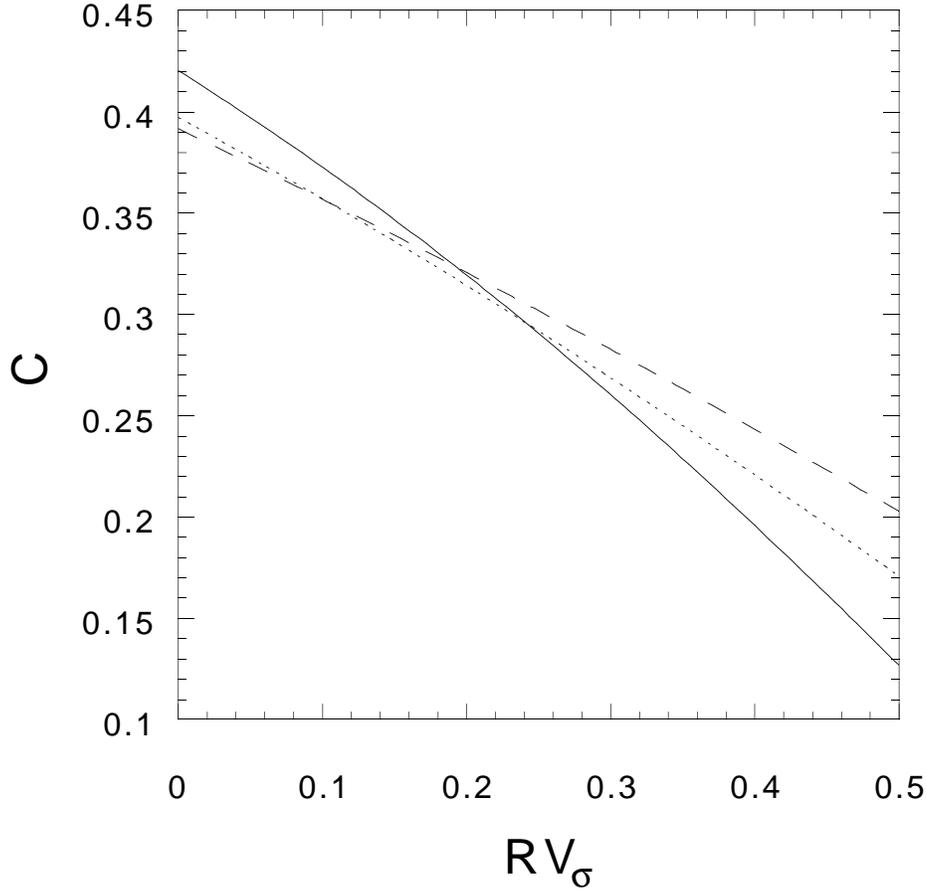

Figure 1: Scalar density factors for various bag radii as a function of $RV_\sigma$ ($m_q = 0$ MeV). The solid, dotted and dashed curves show the results for $R_0 = 0.6, 0.8$ and $1.0 fm$, respectively.

Now we are in a position to present numerical results. First we take $m_q = 0$ MeV, and determine the coupling constants, $g_\sigma^2$ and $g_\omega^2$, so as to fit the binding energy ($-16$ MeV) and the saturation density ($\rho_0 = 0.17 fm^{-3}$) for equilibrium nuclear matter. Furthermore, we choose the $\rho$ meson coupling constant, $g_\rho^2$, so as to reproduce the bulk symmetry energy, 33.2 MeV [11]. The coupling constants and some calculated properties of nuclear matter at saturation density are listed in table 2. Because the nucleon is extended these coupling constants are related to the coupling constants at the meson poles, $g_\sigma(im_\sigma)$, $g_{\omega or \rho}(im_{\omega or \rho})$, by

$$g_\sigma(im_\sigma) = g_\sigma \int d\vec{r} e^{i\vec{k}\cdot\vec{r}} \bar{\psi}_q \psi_q |_{|\vec{k}|=im_\sigma}, \tag{32}$$

$$g_{\omega,\rho}(im_{\omega,\rho}) = g_{\omega,\rho} \int d\vec{r} e^{i\vec{k}\cdot\vec{r}} \psi_q^\dagger \psi_q |_{|\vec{k}|=im_{\omega,\rho}}. \tag{33}$$

Table 2: Coupling constants and calculated properties of equilibrium nuclear matter. The effective nucleon mass, $M_N^\star$, the nuclear compressibility, $K$, and the symmetry energy, $a_4$, are quoted in MeV. The bottom row is for QHD-II. We take $m_\sigma = 550$ MeV, $m_\omega = 783$ MeV and $m_\rho = 770$ MeV.

| $R_0$ | $g_\sigma^2/4\pi$ | $g_\omega^2/4\pi$ | $g_\rho^2/4\pi$ | $M_N^\star$ | $K$ | $a_4$ | $\frac{\delta R}{R_0}$ | $\frac{\delta x}{x_0}$ |
|---|---|---|---|---|---|---|---|---|
| 0.6 | 20.2 | 1.55 | 5.51 | 839 | 220 | 33.2 | -0.03 | -0.07 |
| 0.8 | 22.0 | 1.14 | 5.67 | 851 | 200 | 33.2 | -0.02 | -0.10 |
| 1.0 | 22.5 | 0.95 | 5.69 | 856 | 190 | 33.2 | -0.01 | -0.12 |
| QHD-II | 7.29 | 10.8 | 2.93 | 522 | 540 | 33.6 | — | — |

Using eqs.(32) and (33), the value of the coupling constant at each meson pole is obtained: for example, $g_\sigma^2(im_\sigma)/4\pi = 9.35$, $g_\omega^2(im_\omega)/4\pi = 5.78$ and $g_\rho^2(im_\rho)/4\pi = 27.4$ for $R_0 = 0.8 fm$. It should be pointed out that the strength of the $\omega$ field is considerably less than that required in QHD because the c.m. correction to the bag provides a new source of repulsion [5, 7]. Furthermore, the present model provides values for the nuclear compressibility, $K$, well within the experimental range [12]: $K = 180 \sim 325$ MeV. In the last two columns of table 2 we show the relative modifications (with respect to their values at zero density) of the bag radius, $R$, and the lowest eigenvalue, $x$, at the saturation density. The changes are small (see also Ref.[7]).

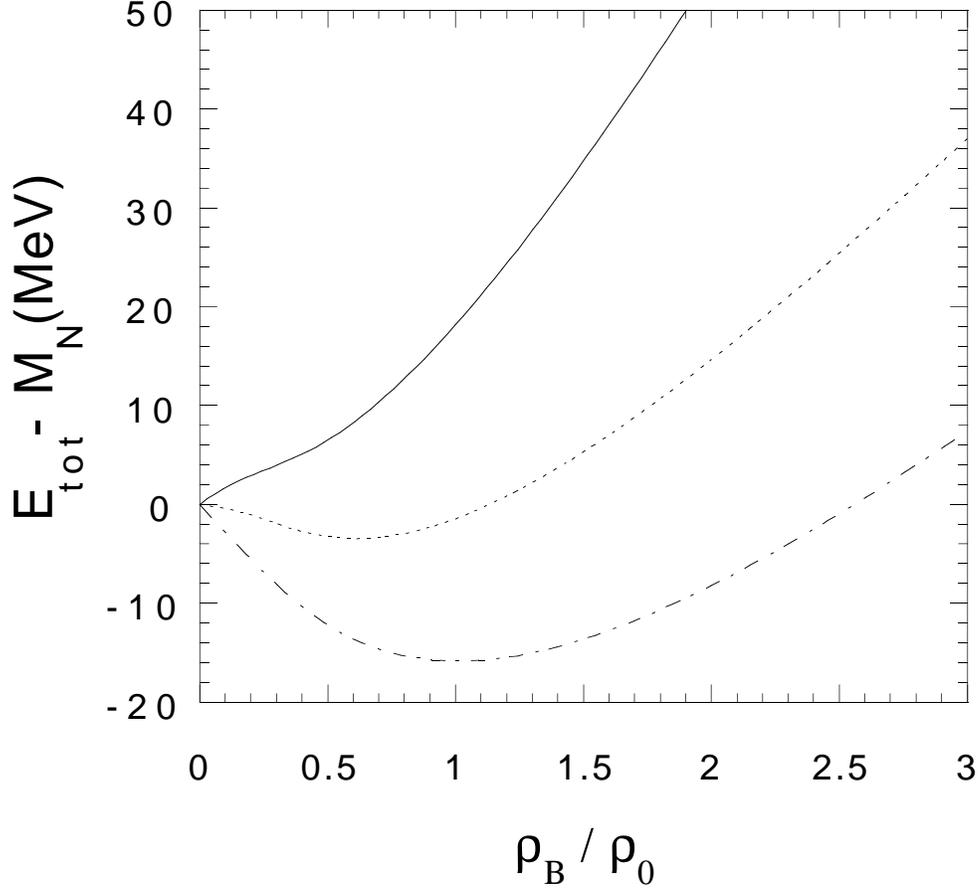

Figure 2: Energy per nucleon for symmetric nuclear matter and for neutron matter. The dot-dashed curve is the saturation curve for nuclear matter with $R_0 = 0.8 fm$. The solid curve (for $g_\rho^2 = 71.2$) and the dotted curve (for $g_\rho^2 = 0$) show the results for neutron matter with the same bag radius.

The total energy per nucleon for both symmetric nuclear matter and for neutron matter (with $R_0 = 0.8 fm$) are presented in fig.2. We see that ignoring the $\rho$ meson coupling yields a shallow bound state of neutron matter around $\rho_B \sim 0.6\rho_0$, but it becomes unbound when the $\rho$ meson contribution is introduced. In fig.3, we show the equation of state (EOS) for neutron matter for both $R_0 = 0.6$ and $1.0 fm$. The $\rho$ meson contribution is again significant in the EOS, while the dependence on the radius of the free nucleon is weak. Although the shape of the EOS for neutron matter in the present model is qualitatively similar to that in QHD-II, it is much softer than in QHD-II because of the lower nuclear compressibility resulting from the internal quark structure of the nucleon.

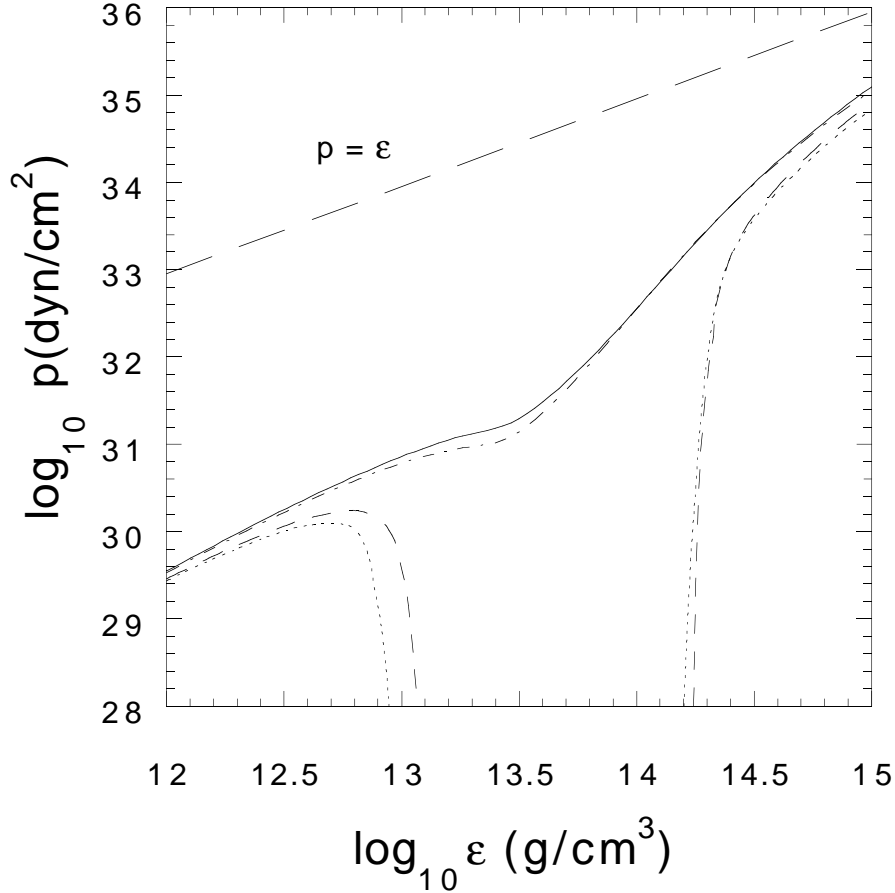

Figure 3: The equation of state for neutron matter. The dashed curve (for $R_0 = 0.6 fm$) and the dotted curve (for $R_0 = 1.0 fm$) show the results when the rho meson contribution is ignored. The solid and dot-dashed curves are the full calculations for $R_0 = 0.6$ and $1.0 fm$, respectively.

Having shown that the QMC model provides an acceptable description of the saturation properties of nuclear matter we now apply it to study the behaviour of the quark condensate in matter. Recent work has linked the quark condensate to a wide range of nuclear phenomena [13] and it is an important parameter in any finite density calculation using QCD sum-rules [14]. From this point we shall deal only with symmetric nuclear matter and introduce a tiny current quark mass for both u and d quarks: for example, $m_q = 10$ MeV. Accordingly, we must readjust the bag parameters and the coupling constants to fit the free nucleon mass and the saturation energy and density. We find that $B$ and $z_0$ become slightly reduced and enhanced, respectively, from those values shown in table 1 and that the coupling constants are $g_\sigma^2/4\pi = 19.1, 20.6, 20.8$ and $g_\omega^2/4\pi = 1.74, 1.32, 1.13$ for $R_0 = 0.6, 0.8, 1.0 fm$, respectively.

The quark condensate at low density can be related to the nucleon $\sigma$ term, $\sigma_N$, which is obtained by using the Hellmann-Feynman theorem [15, 16]. Since we find

$$\left(\frac{dM_N^\star}{dm_q}\right)_R = 3C(\bar{\sigma}), \tag{34}$$

from eqs.(6) and (28), the $\sigma$ term for a nucleon in free space is

$$\sigma_N = 3m_q C(0), \tag{35}$$

where $C(\bar{\sigma})$ is the scalar density factor for a massive quark. We note that it does not change much from that in the massless quark case and that the bag radius dependence is not strong – unlike fig. 1. If we use $C(0) \simeq 0.4$ and $m_q = 10$ MeV, we obtain $\sigma_N \simeq 12$ MeV, which is much smaller than the empirical value [17]: $\sigma_N^{exp} \simeq 45$ MeV. However it should be noted that there are many other potential contributions to $\sigma_N$, such as the meson cloud of the nucleon and its strange quark content [15, 18].

Following Cohen et al. [16], the difference between the quark condensates in matter and vacuum is given by

$$\begin{aligned}Q(\rho_B) - Q(0) &= \frac{1}{2}\left(\frac{d\mathcal{E}}{dm_q}\right), & (36) \\ &\simeq \frac{1}{2}\left(\frac{\partial\mathcal{E}}{\partial M_N^\star}\frac{\partial M_N^\star}{\partial m_q} + \frac{\partial\mathcal{E}}{\partial m_\sigma}\frac{\partial m_\sigma}{\partial m_q} + \frac{\partial\mathcal{E}}{\partial m_\omega}\frac{\partial m_\omega}{\partial m_q}\right), & (37) \\ &\simeq \frac{\sigma_N}{2m_q}\left(\frac{C(\bar{\sigma})}{C(0)}\frac{\partial\mathcal{E}}{\partial M_N^\star} + \chi_\sigma\frac{\partial\mathcal{E}}{\partial m_\sigma} + \chi_\omega\frac{\partial\mathcal{E}}{\partial m_\omega}\right), & (38)\end{aligned}$$

where $\mathcal{E}$ is the total energy density, $\rho_B E_{tot}$, $Q(\rho_B)$ is the in-medium quark condensate, $\langle\rho_B|\bar{q}q|\rho_B\rangle$, and $Q(0)$ that in vacuum. We have used eqs.(34) and (35) and some relations suggested by Cohen et al. [16], $\left(\frac{\partial m_\sigma}{\partial m_q}\right) = \chi_\sigma \left(\frac{\sigma_N}{m_q}\right)$ and $\left(\frac{\partial m_\omega}{\partial m_q}\right) = \chi_\omega \left(\frac{\sigma_N}{m_q}\right)$ with $\chi_\sigma \simeq \left(\frac{m_\sigma}{M_N}\right)$ and $\chi_\omega \simeq \left(\frac{m_\omega}{M_N}\right)$, to derive the last equation. Using eqs.(18) and (30) and the Gell-Mann–Oakes–Renner relation, $2m_q Q(0) = -m_\pi^2 f_\pi^2$, we find

$$\frac{Q(\rho_B)}{Q(0)} \simeq 1 - \frac{\sigma_N \rho_0}{m_\pi^2 f_\pi^2} \left[ \frac{m_\sigma^2}{g_\sigma^2 \rho_0 C(0)}(g_\sigma \bar{\sigma}) + \frac{\chi_\sigma m_\sigma}{g_\sigma^2 \rho_0}(g_\sigma \bar{\sigma})^2 - \frac{\chi_\omega g_\omega^2 \rho_0}{m_\omega^3} \left(\frac{\rho_B}{\rho_0}\right)^2 \right], \qquad (39)$$

where $m_\pi$ is the pion mass (138 MeV) and $f_\pi$ the pion decay constant (93 MeV). Note that the ratio of the quark condensates, $Q(\rho_B)/Q(0)$, is a simple function of the mean $\sigma$ field, $g_\sigma \bar{\sigma}$, and $\rho_B/\rho_0$. In fig.4, the $\sigma$ field values are plotted at low density. We see that the dependence on the radius of the bag is very weak at low density and that the mean field value is well approximated by a linear form for $\rho_B \leq 0.5\rho_0$:

$$g_\sigma \bar{\sigma} = 3V_\sigma \approx 400 \text{ (MeV)} \left(\frac{\rho_B}{\rho_0}\right). \tag{40}$$

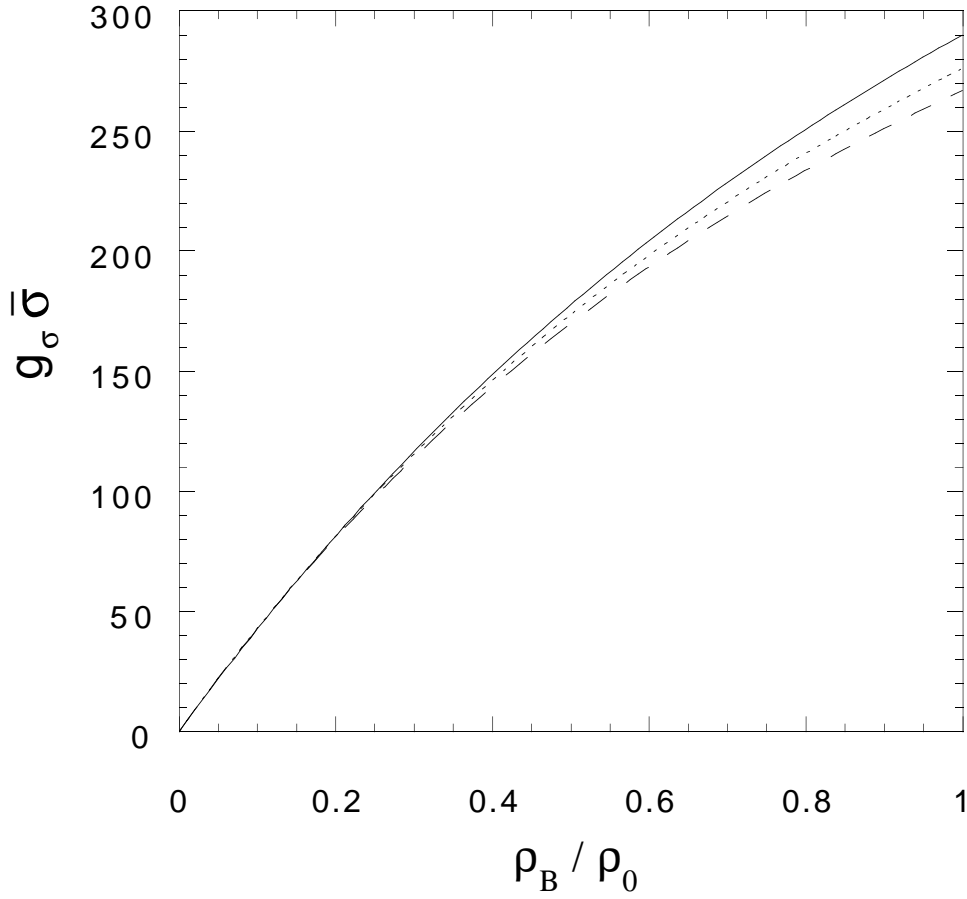

Figure 4: Mean field value of the $\sigma$ meson at low densities (with $m_q = 10$ MeV). The solid, dotted and dashed curves are for $R_0 = 0.6, 0.8$ and $1.0 fm$, respectively.

Hence, if we assume $\sigma_N = 45$ MeV and take $g_\sigma^2 = 250$, $g_\omega^2 = 18$ and $C(0) \simeq 0.4$, we find a simple formula for the ratio at low densities

$$\frac{Q(\rho_B)}{Q(0)} \simeq 1 - 0.331 \left(\frac{\rho_B}{\rho_0}\right) + \mathcal{O}\left(\left(\frac{\rho_B}{\rho_0}\right)^2\right), \tag{41}$$

which is consistent with the model-independent prediction for the ratio of quark condensates proposed by Cohen et al. [16]. At $\rho_B = 0.5\rho_0$, the in-medium quark condensate is reduced by roughly 15% from its value in vacuum.

Furthermore, since the effective nucleon mass can be expanded at low densities as

$$M_N^\star(m_q^\star) \simeq M_N(m_q) - V_\sigma \times \left.\frac{dM_N^\star}{dm_q^\star}\right|_{m_q^\star = m_q}, \tag{42}$$

$$= M_N - C(0)(g_\sigma \bar{\sigma}), \tag{43}$$

we obtain a relation between the quark and nucleon masses

$$\frac{M_N - M_N^\star}{m_q - m_q^\star} \simeq 3C(0) \left(= \left.\frac{dM_N^\star}{dm_q^\star}\right|_{m_q^\star = m_q}\right), \tag{44}$$

where we have used eq.(6). This relation is satisfied within a few percent up to $\rho_B/\rho_0 \sim 0.5$. If we now solve eq.(43) for $M_N^\star/M_N$, we find

$$\frac{M_N^\star}{M_N} \simeq 1 - \frac{C(0)}{M_N}(g_\sigma \bar{\sigma}), \tag{45}$$

and, using eq.(40) and $C(0) \simeq 0.4$, we find

$$\frac{M_N^\star}{M_N} \simeq 1 - 0.170 \left(\frac{\rho_B}{\rho_0}\right). \tag{46}$$

Comparing this with eq.(41) we see that $Q(\rho_B)/Q(0)$ is essentially $(M_N^\star/M_N)^2$ at low density. This result is intermediate between the cubic dependence found by Brown and Rho [19] and the linear dependence found by Cohen et al. [16]. If on the other hand we were to take the limit of heavy quarks considered earlier it is easy to show that one recovers the linear dependence.

Finally, we would like to add some caveats concerning the present model: (1) The basic idea is that the mesons are locally coupled to the quarks. This is certainly common for pions in models like the chiral or cloudy bag [20], but may be less justified for vector mesons which are not collective states. (2) The model is best suited to low and moderate density regions – certainly no higher than $\rho_B \sim 3.0\rho_0$ – because short-range q-q correlations are not taken into account. At high densities these would be expected to dominate. (3) The pionic cloud of the nucleon [20] should be considered explicitly in any truly quantitative study of the properties of the nucleon in-medium.

In summary, we have presented the quark-meson coupling model based on a mean field description of nucleon bags interacting through the exchange of $\sigma$, $\omega$ and $\rho$ mesons, and shown that it can provide an excellent description of the properties of both symmetric and asymmetric nuclear matter. The relationship to QHD, and particularly the role of the internal structure of the nucleon, was clarified. Finally, the model was shown to provide a reasonable description of the quark condensate in-medium. The present model seems to be the simplest possible extension of QHD to incorporate explicit quark degrees of freedom.

This work was supported by the Australian Research Council.